# The stability, electronic structure, and optical property of TiO$_2$ polymorphs

Tong Zhu and Shang-Peng Gao[a)]

Department of Materials Science, Fudan University, Shanghai 200433, P. R. China

Phonon density of states calculation shows that a new TiO$_2$ polymorph with tridymite structure is mechanically stable. Enthalpies of 9 TiO$_2$ polymorphs under different pressure are presented to study the relative stability of the TiO$_2$ polymorphs. Band structures for the TiO$_2$ polymorphs are calculated by density functional theory with generalized gradient approximation and the band energies at high symmetry $\boldsymbol{k}$-points are corrected using the GW method to accurately determine the band gap. The differences between direct band gap energies and indirect band gap energies are very small for rutile, columbite and baddeleyite TiO$_2$, indicating a quasi-direct band gap character. The band gap energies of baddeleyite (quasi-direct) and brookite (direct) TiO$_2$ are close to that of anatase (indirect) TiO$_2$. The band gap of the newly predicted tridymite-structured TiO$_2$ is wider than the other 8 polymorphs. For optical response calculations, two-particle effects have been included by solving the Bethe-Salpeter equation for Coulomb correlated electron-hole pairs. TiO$_2$ with cotunnite, pyrite, and fluorite structures have optical transitions in the visible light region.

## I. INTRODUCTION

Even after half a century of research,[1,2] investigation of the fundamental properties of TiO$_2$ crystal phases remains very important properly due to their important role to effectively utilize solar energy. For instance, photocatalytic splitting of water into H$_2$ and O$_2$,[3] photovoltaic generation of electricity,[4] degradation of environmentally hazard materials,[5,6] and reduction of CO$_2$ into hydrocarbon fuels.[7] The gap between valence and conduction bands and the optical absorption property are vital to all these applications.

Titanium dioxide exists in many polymorphs. Among them, anatase, rutile, and brookite[8] are well known minerals in nature. Furthermore, TiO$_2$ has a rich phase diagram at elevated pressure. High pressure x-ray-diffraction[9–14] and the Raman spectroscopy[15–20] studies have proved that rutile and anatase can be transformed to a columbite ($\alpha$-PbO$_2$) phase at high pressure. Recent x-ray-diffraction studies reported that the columbite phase is only formed at about 7 GPa during decompression from a higher pressure phase,[12–14] while the transformation of rutile and anatase directly to columbite-structured TiO$_2$ has been observed at 5 GPa[15–20] in Raman studies. Columbite-structured TiO$_2$ was also discovered in the suevite from the Ries crater in Germany[21]. It has been found that columbite-structured TiO$_2$ are transformed to a baddeleyite structure between 12 GPa and 17 GPa in x-ray-diffraction[12,22] and Raman[17,19]

---

[a)]Electronic mail: gaosp@fudan.edu.cn

studies. However, calculations indicated the transition pressure to be 26 GPa[23] or 31 GPa[24]. A TiO$_2$ polymorph with cotunnite structure was observed at pressure higher than 37.4 GPa.[25-27] A number of observations suggested a transformation to a cubic phase at about 60 GPa but without sufficient data available to fully determine the structure.[19,28,29] However, because it has been known that several rutile-structured metal oxides at 0 GPa can be transformed to fluorite-structured phases at high pressure and Rietveld refinement of x-ray-diffraction data from three rutile-structured oxides (SnO$_2$, PbO$_2$, and RuO$_2$) revealed that high-pressure phase in these systems actually adopts a pyrite structure,[30] it has been postulated that this cubic phase of TiO$_2$ adopts a fluorite or pyrite structure. Whether or not there are other possible structures is still an open question.

For the electronic and optical properties of TiO$_2$ polymorphs, the experimental studies are mainly about the mineral phases. The electronic band structure of rutile has reported values of 3.3±0.5 eV[31] (photoemission spectroscopy (PES) and inverse photoemission spectroscopy (IPES)), 3.6±0.2 eV[32] (PES and IPES for rutile (110) surface). Hardman et al. reported the measurement of valence-band structure of rutile along Γ-Δ-X and Γ-Σ-M directions by PES[33]. For anatase and brookite, there are no reported measurements of the electronic band gap from combined PES and IPES measurements. The optical band gap is reported at ~3.0 eV for rutile[34,35], ~3.4 eV for anatase[36] and ~3.3 eV for brookite[37]. The indirect absorption edge has been measured to be 3.0 eV for rutile[35,38], 3.2 eV for anatase[36,39–42]. The direct exciton is reported at 3.57 eV for rutile[34,43,44] and 3.68~3.90 eV for anatase[45–47] according to the results of absorption, photoluminescence, and Raman-scattering technique.

It has been generally observed for semiconductors and insulators, the band gap is underestimated in the density functional theory (DFT) calculations with local density approximation (LDA) or generalized gradient approximation (GGA) for the exchange correlation functional.[48,49] Ab initio many-body perturbation theory with GW approximation is regarded as an accurate way to predict the band structure.[48,50–53] Available band gap data calculated by the GW method in literatures for TiO$_2$ polymorphs are listed in TABLE I.[8,54–62] There are still no theoretical band structures for Columbite and Baddeleyite phases of TiO$_2$. Mattesini et al. demonstrated the possible optical transition in visible light region for pyrite and fluorite TiO$_2$ from the DFT calculation.[61] GW method[62] and quantum Monte Carlo[63] studies show that fluorite TiO$_2$ has band gap energy lying in the visible light region, though the GW calculation for pyrite is still absent.

Optical absorption process accompanies with the generation of electron-hole pairs. An accurate way to treat the electron-hole interaction is solving the Bethe-Salpeter equation (BSE) for Coulomb correlated electron-hole pairs. Kang and Hybertson[55] as well as Chiodo et al.[56] investigated the optical excitation energies of rutile and anatase TiO$_2$.

Lawler et al. investigated the birefringence of rutile and anatase phase TiO$_2$[64]. Landmann et al. have calculated the three natural occurring TiO$_2$ polymorphs rutile, anatase, and brookite[58]. MBPT calculations of band structure and optical property for newly proposed structures, such as columbite, baddeleyite, cotunnite, and pyrite structures, are still lacked (Table II).

In this paper, a new possible TiO$_2$ polymorph with tridymite structure is proposed. For the convenience of discussion, we classified TiO$_2$ polymorphs in three categories: phases that have been found in minerals: the well-known rutile, anatase and brookite, and the less well-known columbite-structured TiO$_2$ which has been found in shocked garnet gnersses, high pressure phases that have been reported in literature (baddeleyite, cotunnite, pyrite, and fluorite), and a new tridymite-structured TiO$_2$ phase we proposed in this paper. The structure optimization and electronic band structure calculations are carried out for those 9 TiO$_2$ polymorphs. GW method is adopted to calculate the band energies at high-symmetry $k$-points in the first Brillouin zone. BSE method is used to calculate the optical absorption spectrum. The GW band gap energies and the optical absorption spectrum based on BSE method would be helpful for future research work on electrical, optical, and transport properties of these TiO$_2$ polymorphs.

## II. COMPUTATIONAL METHODS

A planewave pseudopotential code ABINIT[53,65,66] is employed for both DFT and MBPT calculations. Structure and ground state electronic structure calculations are based on the DFT-GGA. The Perdew-Burke-Ernzerhof GGA functional is used to describe the exchange-correlation potential.[67] All pseudopotentials in this work are generated using the OPIUM package[68] in the Troullier-Martins Scheme[69]. For the pseudopotential of the Ti, it has been reported by Kang and Hybertsen[55] and also confirmed by our test that dividing the n=3 shell of Ti into frozen core (3s, 3p) and valence (3d) contributions introduces a significant error to the band gap energy. Therefore, all three semicore sub-shells of Ti, namely, 3s, 3p, and 3d are treated as valence electrons. The plane wave cutoff energy is set to 1633 eV (60 Hartree) in the DFT calculation. The Monkhorst-Pack scheme is used for the $k$-point sampling in the first Brillouin zone. A 6×6×8 $k$-point grid is chosen for rutile. A 6×6×3 $k$-point grid is chosen for anatase. A 4×4×4 $k$-point grid is chosen for TiO$_2$ polymorphs with brookite, columbite, baddeleyite, cotunnite, and pyrite structures. An 8×8×8 $k$-point grid is chosen for fluorite-structured TiO$_2$ with primitive cell adopted in the calculation. A 6×6×4 $k$-point grid is chosen for tridymite phase. Structural optimization is carried out using the Broyden-Fletcher-Goldfarb-Shanno minimization (BFGS).[70] In order to be consistent with the full *ab initio* philosophy of this work, all the results shown here have been obtained using the calculated geometries.

Band energies at high symmetry $k$-points are corrected using the standard one-shot $G_0W_0$ method.[48,52,53] Following the standard approach, Kohn-Sham eigenvalues and eigenfunctions will be firstly obtained by DFT-GGA calculation and then used as a starting point to do the GW correction.[48,52,53] In order to include all the high symmetry $k$-points and consider the calculation efficiency, 4×4×4 and 6×6×2 $k$-points are both chosen for tridymite-structured $TiO_2$. The screening in the GW calculation is treated with the plasmon pole approximation.[48,53] The dielectric matrix is evaluated at an imaginary frequency of 13.6 eV for all $TiO_2$ polymorphs. The polarization function, which is necessary to evaluate the screened interaction, is calculated within the random phase approximation. The numbers of bands used to calculate the screening and the self-energy in the GW method are chosen to be 344 for rutile, 368 for anatase, baddeleyite-structured $TiO_2$, and pyrite-structured $TiO_2$, 512 for brookite, 384 for columbite-structured $TiO_2$, 432 for cotunnite-structured $TiO_2$, 332 for fluorite-structured $TiO_2$, and 320 for tridymite-structured $TiO_2$. The cut-off energy of the planewave is set to 490 eV (18 Hartree) to represent the independent particle susceptibility, the dielectric matrix and to generate the exchange part of the self-energy operator.

In order to get the excitation properties, such as the absorption of light, the interaction between the excited electron and the hole is included by solving the Bethe-Salpeter equation.[71–74] The Tamm-Dancoff approximation that neglects the two off-diagonal coupling part in the two-particle Hamiltonian is applied. The Haydock iterative technique is used to invert the Bethe-Salpeter equation, so that we can make use of the functions in the W contribution to the kernel in order to make the calculation less cumbersome. The coulomb term of the BS Hamiltonian is evaluated using the truly non-local screening function W. In the BSE calculation, we use the Kohn-Sham eigenvalues and wave functions to construct the transition space. To consider the effect of the self-energy correction, the Kohn-Sham energies are corrected by a scissors operator with its energy given by the difference between the band gap calculated with DFT-GGA and GW methods. This permits us to avoid a cumbersome GW calculation for each state included in the transition space. A complex shift of 0.15 eV is used to avoid divergences in the expression for the macroscopic dielectric function.

The phonon density of states (DOS) of tridymite-structured $TiO_2$ are calculated by DFT-GGA using the CASTEP code[75] within Materials Studio 5.5 Package. Norm-conserving pseudopotential is required for linear response calculations or finite displacement calculations that require LO-TO splitting correction using the CASTEP code. We want to emphasize that the CASTEP code is only used for phonon DOS study in this work and this do not affect the consistency of the structural optimization, pressure induced phase transformation, band structure and optical spectra that are calculated using the ABINIT code.

# III. RESULTS AND DISCUSSION

## A. Structure and relative stability of TiO$_2$ polymorphs

The polyhedral structure of TiO$_2$ polymorphs investigated in this work are given in Fig. 1. By searching possible structures of AB$_2$ types, we proposed a new TiO$_2$ polymorphs with tridymite (a high-temperature polymorph of quartz) structure. Phonon DOS of tridymite-structured TiO$_2$ are displayed in Fig. 2 and all the phonon modes have positive (real) frequencies, indicating that tridymite-structured TiO$_2$ can be mechanically stable. The tridymite-structured TiO$_2$ belongs to space group P6$_3$/mmc. Each Ti ion is tetrahedrally coordinated to 4 O ions.

Geometry optimized structures at zero pressure for the 9 TiO$_2$ polymorphs are listed in Table II. For the natural phases, each Ti ion is octahedrally coordinated to six O ions. As shown in TABLE II, the lattice constants calculated by DFT-GGA for rutile, anatase, brookite, and columbite-structured TiO$_2$ agree well with other experimental results [12,76-79]. The calculated coordinate of O atom occupying the 4f Wyckoff position in rutile is (0.305, 0.305, 0), which is same as the experimental results [76,77]. The calculated internal coordinate of O atom occupying the 8e Wyckoff position in anatase is (0, 0, 0.206) that is close to the experimental reported value (0, 0, 0.208) [76,78].

In order to compare with experimental lattice parameters (a=4.64 Å, b=4.76 Å, c=4.81 Å, and β=99.2°) which are measured at 20.3 GPa, we also carried out the structure optimization under 20.3 GPa for TiO$_2$ that gives lattice parameters (a=4.643 Å, b=4.865 Å, c=4.811 Å, and β=98.83°). For cotunnite-structured TiO$_2$, calculated lattice constants at 60 GPa are closer to the experimental results reported by Nishio-Hamane et al. (a=5.028 Å, b=2.930 Å, c=5.889 Å at 59.6 GPa)[25] than that measured by Dubrovinsky et al. (a=5.163 Å, b=2.599 Å, c=5.966 Å at 60 GPa)[27]. There are no experimental data for fluorite and pyrite phases. Muscat et al.[24] and Mattesini et al.[61] have reported theoretical lattice parameters using DFT-GGA and DFT-LDA respectively. For consistency, theoretical structure data shown in Table II are used in the band structure and optical property calculation in this paper.

Enthalpy of the TiO$_2$ polymorphs under hydrostatic pressure from 0 GPa to 60 GPa are shown in Fig. 3. At zero pressure, it can be found that the energy (enthalpy) of columbite structured TiO$_2$ is lower than that of rutile and higher than that of brookite and anatase. Anatase can transformed to brookite at 3.2 GPa and to rutile at pressure higher than 5.6 GPa. Experiments has observed that anatase may either transform directly to rutile, or initially to brookite and then to rutile.[80,81] The phase transformation from anatase structure to columbite structure occurs at about 3.5 GPa, and the transformation from the columbite structure to baddeleyite structure occurs at about 12 GPa, that is in reasonable agreement with experimental observations that anatase structure transforms to the columbite structure at 5 GPa[15–20] and the columbite structure transforms to the baddeleyite structure between 12 and 17 GPa.[12,17,19,22] The phase transformation from the baddeleyite structure to cotunnite structure occurs at about 38 GPa, agreeing with the experiment that cotunnite phase appeared at 37.4 GPa[25]. The cotunnite phase is the most stable phase above 37.4 GPa to at least 60 GPa, in agreement with the experimental observation[25]. The pyrite-structured $TiO_2$ become more stable than the fluorite structure above 22 GPa and become more stable than the rutile at pressure above 45 GPa. At zero pressure, the enthalpy (energy) of tridymite-structured $TiO_2$ lower than rutile, but at elevated pressure tridymite-structured $TiO_2$ become less stable. This indicates that the newly proposed tridymite-structured $TiO_2$ cannot be obtained from high pressure phase transformation from the nature minerals but may be synthesized via chemical reactions under ambient pressure.

B. Electronic band structure and density of states

The band energies at high symmetry $k$-points from both DFT-GGA and GW calculations are listed in TABLE III. Theoretical lattice parameters and atomic positions in TABLE II are used. For easy comparison, the band gaps calculated by DFT-GGA and GW methods are listed in TABLE IV. First, we look at the electronic band structures and DOS for the 4 mineral phases (Fig. 4). DFT-GGA band structure of rutile (Fig.4(a)) indicates that it has a direct band gap 1.80 eV at Γ point. But with GW correction, rutile has an indirect band gap of 3.18 eV with the valence band maximum (VBM) at Γ and conduction band minimum (CBM) at R. The lowest conduction band energy at R is slightly lower than that at Γ by about 0.05 eV. Kang and Hybertsen[55] have found the same phenomenon and reported an indirect GW band gap of 3.34 eV and a direct band gap of 3.38 eV at Γ. The direct GW band gap of 3.23 eV at Γ from our calculation agrees well with the experimental result 3.3±0.5 eV measured by PES and IPES[31]. DFT-GGA calculation shows that anatase has an indirect band gap with the VBM at Δ (0.4318,0.4318,0) between M (0.5,0.5,0) and Γ and the CBM at Γ(Fig.4(b)). The indirect DFT-GGA band gap 2.08 eV agrees well with the results given by Mo et al. (2.04 eV)[8] and Labat et al. (2.08 eV)[54] from DFT calculations. The indirect GW band gap from M to Γ is 3.71 eV, agreeing well with 3.73 eV calculated by Landmann et al.[58]. A minimum band gap of 3.64 eV can be deduced approximately from GW band energies at M and Γ and the DFT band dispersion from M and Δ. Brookite has a direct band gap at Γ (Fig. 4(c)) and its GW band gap (3.86 eV) is close to that of anatase (3.71 eV) whereas anatase has an indirect band gap character. Our calculation shows that columbite-structured $TiO_2$ has an indirect band gap with the VBM at Γ and CBM at Z (Fig.5(a)) From the GW calculation, the indirect band gap for columbite is 4.09 eV and the minimum direct band gap at Γ is 4.15 eV. The direct band gap is only 60 meV larger than the indirect band gap. There are still no electronic band structure data reported for columbite-structured $TiO_2$ in literature.

TiO$_2$ polymorphs with baddeleyite, cotunnite, pyrite, and fluorite structures have been studied in literatures as high pressure phases. The DFT-GGA band structure calculation for baddeleyite TiO$_2$ shows that the lowest conduction band energies at Γ, D, and D are very close. GW calculation indicates that the minimum band gap of the baddeleyite TiO$_2$ is indirect with the VBM at Γ and CBM at B (Table II). The GW band gap of the baddeleyite TiO$_2$ is 3.69 eV, which is very close to that of anatase (3.71 eV). Cotunnite structure has a GW band gap of 2.89 eV, lying in the short-wavelength range of visible light.

Fig.5 (c) and Fig. 5(d) show band structure of two TiO$_2$ polymorphs with cubic lattice: pyrite-structured TiO$_2$ and fluorite-structured TiO$_2$. Pyrite has an indirect band gap with VBM at Γ and CBM at R and its DFT-GGA band gap of 1.39 eV is close to the DFT-LDA band gap of 1.44 eV reported by Mattesini et al.[61] The band gap calculated by the GW method is 2.55 eV for pyrite TiO$_2$. DFT-GGA calculation shows fluorite TiO$_2$ has an indirect band gap of 1.09 eV with the VBM at X and CBM at Γ (Fig.5(d)), comparable with the DFT-LDA value of 1.44 eV[61] and DFT-GGA value of 1.04 eV.[62] The indirect band gap from X to Γ for fluorite TiO$_2$ from GW calculations is 2.26 eV, about 0.3 eV smaller than the pyrite TiO$_2$. Among the 9 TiO$_2$ polymorphs studied, the band gap of TiO$_2$ with cotunnite, pyrite, and fluorite structures meet the requirement of photocatalyst candidate with visible light catalytic activity: 2~3.1 eV[82].

The tridymite-structured TiO$_2$ has a direct band gap at Γ (Fig. 6(a)). The maximum valence band energies at Γ is very close to that at A and M points. The band gaps calculated by the GW method is 5.67 eV respectively, larger than the other TiO$_2$ polymorphs studied in this paper.

C. Optical-absorption properties

Real and imaginary parts of the frequency-dependent macroscopic dielectric function ε(ω) calculated by the Bethe-Salpeter equation for 4 mineral TiO$_2$ phases (rutile, anatase, brookite, and columbite-structured TiO$_2$) are presented in Fig. 7(a)~(d). The dielectric functions for rutile and anatase, which are tetragonal structure, are resolved into two components: the in plane component **E**⊥*c* and the out-of-plane component **E**//*c* to study their optical anisotropy. Three dielectric components parallel to *a*, *b*, and *c* axis are resolved for the brookite and columbite structured TiO$_2$ that have orthorhombic crystal lattice. Compared with experimental spectra obtained by means of spectroscopic elipsometry for the imaginary of the dielectric function of rutile[83], the peaks below the 6 eV from the calculated results agree well with the experimental data both for the location and the amplitude whereas the peaks above 6 eV is red shift slightly (0.2 eV- 0.4 eV) and the amplitude is higher comparing to the experimental data. For the real part of the dielectric function, the value of Re $\varepsilon(\omega = 0)$ is very close with the experimental data. This justifies the reliability of the macroscopic static dielectric constants in TABLE V obtained from the Re $\varepsilon(\omega = 0)$. For anatase, main features in the experimental spectra[84]. For columbite-structured $TiO_2$, a very sharp peak appears on the absorption threshold of the imaginary dielectric function for **E**//***a***.

As many optical experiments actually probe the absorption coefficients α, the normal-incidence absorption coefficients are also calculated and compared with the experimental results from Kramers-Kronig analysis of absorption coefficient data[37,38] (Fig. 8). The agreement between experimental and calculated coefficients of rutile is very good for **E**⊥***c*** component and is less satisfactory for **E**//***c*** component. The agreement near the absorption threshold is poor due to a constant value (0.15 eV) giving the complex shift to avoid divergences in the continued fraction in the iterative Haydock technique to calculate the macroscopic dielectric function and this value mimics the experimental broadening of the absorption peaks.

The calculated imaginary part of dielectric function (absorption spectrum) for $TiO_2$ polymorphs with baddeleyite, cotunnite, pyrite and fluorite structures are shown in Fig. 9. There are still no relevant experimental spectra available for these phases found in high pressure studies. Baddeleyite structure has a monoclinic unit cell and cotunnite structure has an orthorhombic unit cell. Three components along ***a***, ***b***, ***c*** cell vector directions are given separately. For baddeleyite, the onset of **E**//***a*** and **E**//***b*** component is very close and that of the **E**//***c*** component is higher. It can be found that cotunnited-structured, pyrite-structured, and fluorite-structured $TiO_2$ show optical absorption in the visible light range. Pyrite-structured $TiO_2$ has a first absorption peak at 3.1 eV (peak A) and fluorite-structured $TiO_2$ has a prominent peak A at 3.0 eV which is separated from leading features B and C.

Imaginary part of dielectric functions (absorption spectra) for tridymite-structured $TiO_2$ are shown in Fig. 6(b). The optical absorption spectra can serve as good reference for structure characterization. The absorption threshold energy for tridymite-structured $TiO_2$ are higher than that of the other $TiO_2$ polymorphs shown in Figs. 7 and 9 and the spectra feature differences can be easily recognized. Tridymite structure has a hexagonal lattice, anisotropic optical absorption can be identified from the in plane (**E**⊥***c***) and out-of-plane (**E**//***c***) components.

## IV. CONCLUSIONS

By searching the equilibrium structure and calculating their phonon DOS, a new possible mechanically stable $TiO_2$ polymorph with tridymite structure, has been found. Enthalpies of $TiO_2$ polymorphs show that tridymite-structured $TiO_2$ has lower enthalpy than rutile at pressure lower than 170 MPa and becomes less stable at higher pressure. So the newly proposed tridymite-structured $TiO_2$ cannot be found in high pressure studies but might be synthesized by chemical reaction at ambient pressure.

Band gap calculations using the GW method show that the band gaps of the newly found tridymite-structured $TiO_2$ polymorph (5.67 eV) is wider than that of the other 8 $TiO_2$ polymorphs studied in this work: 3.71 eV for anatase, 3.25 eV for rutile, 3.86 eV for brookite, 4.09 eV for columbite-structured $TiO_2$, 3.69 eV for baddeleyite-structured $TiO_2$, 2.89 eV for cotunnite-structured $TiO_2$, 2.55 eV for pyrite-structured $TiO_2$, 2.26 eV for fluorite-structured $TiO_2$. Among the 9 $TiO_2$ polymorphs, brookite, and tridymite-structured $TiO_2$ have direct band gap. For rutile, columbite-structured, and baddeleyite-structured $TiO_2$, the direct band gap energies at $\Gamma$ are very close to the indirect band gap energies, indicating a quasi-direct band gap character. The band gap of brookite and baddeleyite-structured $TiO_2$ is very close to that of anatase. Optical absorption properties are analyzed based on the macroscopic dielectric function calculation by solving the two-particle function BSE. We have shown that $TiO_2$ polymorphs with cotunnite, pyrite, and fluorite structures have optical transitions in the region of the visible light. The GW band gap energies and optical absorption spectra present in this article can serve as a guide in the promising applications for the $TiO_2$ polymorphs.

ACKNOWLEDGMENTS
This work is supported by the State Key Development Program of Basic Research of China (Grant no. 2011CB606406). The computational resources utilized in this research are provided by Shanghai Supercomputer Center and we would like to thank Dr. Tao Wang for his support in using the supercomputer facility.

[1] F. A. Grant, Rev. Mod. Phys. **31**, 646 (1959).

[2] R. G. Breckenridge and W. R. Hosler, Phys. Rev. **91**, 793 (1953).

[3] A. Fujishima and K. Honda, Nature **238**, 37 (1972).

[4] B. Oregan and M. Gratzel, Nature **353**, 737 (1991).

[5] M. A. Fox and M. T. Dulay, Chem. Rev. **93**, 341 (1993).

[6] M. R. Hoffmann, S. T. Martin, W. Choi, and D.W. Bahnemann, Chem. Rev. **95**, 69 (1995).

[7] V. P. Indrakanti, J. D. Kubicki, and H. H. Schobert, Energy Environ. Sci. **2**, 745 (2009).

[8] S.-D. Mo and W. Y. Ching, Phys. Rev. B **51**, 13023 (1995).

[9] N. A. Bendelia, S. V. Popova, and L. F. Vereshch, Geochem. Int. **3**, 387 (1966).

[10] R. G. Mcqueen, J. C. Jamieson, and S. P. Marsh, Science **155**, 1401 (1967).

[11] H. Julian and J. M. Lger, Physica B **192**, 233 (1993).

[12] L. Gerward and J. Olsen, J. Appl. Crystallogr. **30**, 259 (1997).

[13] J. S. Olsen, L. Gerward, and J. Z. Jiang, J. Phys. Chem. Solids **60**, 229 (1999).

[14] T. Arlt, M. Bermejo, M. A. Blanco, L. Gerward, J. Z. Jiang, J. Staun Olsen, and J. M. Recio, Phys. Rev. B **61**, 14414 (2000).

[15] J. F. Mammone, S. K. Sharma, and M. Nicol, Solid State Commun. **34**, 799 (1980).

[16] J. F. Mammone, M. Nicol, and S. K. Sharma, J. Phys. Chem. Solids **42**, 379 (1981).

[17] H. Arashi, J. Phys. Chem. Solids **53**, 355 (1992).

[18] L. G. Liu and T. P. Mernagh, Eur. J. Mineral. **4**, 45 (1992).

[19]K. Lagarec and S. Desgreniers, Solid State Commun. **94**, 519 (1995).

[20]T. Sekiya, S. Ohta, S. Kamei, M. Hanakawa, and S. Kurita, J. Phys. Chem. Solids **62**, 717 (2001).

[21]A. E. Goresy, M. Chen, P. Gillet, L. Dubrovinsky, G. Graup, R. Ahuja, Earth Planet. Sci. Lett. **192**, 485 (2001).

[22]H. Sato, S. Endo, M. Sugiyama, T. Kikegawa, O. Shimomura, and K. Kusaba, Science **251**, 786 (1991).

[23]T. Sasaki, J. Phys.: Condens. Matter **14**, 10557 (2002).

[24]J. Muscat, V. Swamy, and N. M. Harrison, Phys. Rev. B **65**, 224112 (2002).

[25]D. Nishio-Hamane, A. Shimizu, R. Nakahira, K. Niwa, A. Sano-Furukawa, T. Okada, T. Yagi, and T. Kikegawa, Phys. Chem. Minerals, **37**, 129 (2010)

[26]L. S. Dubrovinsky, N. A. Dubrovinskaia, V. Swamy, J. Muscat, N. M. Harrison, R. Ahuja, B. Holm, B. Johansson, Nature, **410**, 653 (2000)

[27]M. Mattesini, J. S. de Almeida, L. Dubrovinsky, N. Dubrovinskaia, B. Johansson, and R. Ahuja, Phys. Rev. B **70**, 212101 (2004).

[28]S. Endo, I. Takenaka, and H. Arashi, AIRAPT Conf. Proc. **309**, 371 (1979).

[29]Y. Syono, K. Kusaba, M. Kikuchi, and K. Fukuoka, Geophys. Monogr. **39**, 385 (1987).

[30]J. Haines, J. M. Lger, and O. Schulte, Science **271**, 629 (1996).

[31]Y. Tezuka, S. Shin, T. Lshii, T. Ejima, S. Suzuki, and S. Sato, J. Phys. Soc. Jpn. **63**, 347 (1994).

[32]S. Rangan, S. Katalinic, R. Thorpe, R. A. Bartynski, J. Rochford, and E. Galoppini, J. Phys. Chem. C **114**, 1139 (2010).

[33]P. J. Hardman, G. N. Raikar, C. A. Muryn, G. van der Laan, P. L. Wincott, G. Thornton, D. W. Bullett, and P. A. D. M. A. Dale, Phys. Rev. B **49**, 7170 (1994).

[34]J. Pascual, J. Camassel, and H. Mathieu, Phys. Rev. B **18**, 5606 (1978).

[35]A. Amtout and R. Leonelli, Phys. Rev. B **51**, 6842 (1995).

[36]H. Tang, F. Lévy, H. Berger, and P. E. Schmid, Phys. Rev. B **52**, 7771 (1995).

[37]A. Mattsson and L. Osterlund, J. Phys. Chem. C **114**, 14121 (2010).

[38]M. Cardona and G. Harbeke, Phys. Rev. **137**, A1467 (1965).

[39]U. Diebold, Surf. Sci. Rep. **48**, 53 (2003).

[40]H. Tang, K. Prasad, R. Sanjinès, P. E. Schmid, and F. Lévy, J. Appl. Phys. **75**, 2042 (1994).

[41]L. Kavan, M. Grtzel, S. E. Gilbert, C. Klemenz, and H. J. Scheel, J. Am. Chem. Soc. **118**, 6716 (1996).

[42]H. Tang, H. Berger, P. E. Schmid, F. Lvy, and G. Burri, Solid State Commun. **87**, 847 (1993).

[43]K. Vos, J. Phys. C **10**, 3917 (1977).

[44]F. M. F. de Groot, J. Faber, J. J. M. Michiels, M. T. Czyżyk, M. Abbate, and J. C. Fuggle, Phys. Rev. B **48**, 2074 (1993).

[45]Z. Wang, U. Helmersson, and P. O. Kall, Thin Solid Films **405**, 50 (2002).

[46]B. Liu, L. Wen, and X. Zhao, Mater. Chem. Phys. **106**, 350 (2007).

[47]M. M. Hasan, A. S. M. A. Haseeb, R. Saidur, and H. H. Masjuki, Int. J. Chem. Biomolecular Eng. **1**, 92 (2008).

[48]R. W. Godby, M. Schlüter, and L. J. Sham, Phys. Rev. B **37**, 10159 (1988).

[49]R. O. Jones and O. Gunnarsson, Rev. Mod. Phys. **61**, 689 (1989).

[50]F. Aryasetiawan and O. Gunnarsson, Rep. Prog. Phys. **61**, 237 (1998).

[51]G. Onida, L. Reining, and A. Rubio, Rev. Mod. Phys. **74**, 601 (2002).

[52]M. S. Hybertsen and S. G. Louie, Phys. Rev. B **34**, 5390 (1986).

[53]X. Gonze et al., Comput. Phys. Commun. **180**, 2582 (2009).

[54]F. Labat, P. Baranek, C. Domain, C. Minot, and C. Adamo, J. Chem. Phys. **126**, 154703 (2007).

[55]W. Kang and M. S. Hybertsen, Phys. Rev. B 82, 085203 (2010).

[56]L. Chiodo, J. M. Garcia-Lastra, A. Iacomino, S. Ossicini, J. Zhao, H. Petek, and A. Rubio, Phys. Rev. B **82**, 045207 (2010).

[57]C. E. Patrick and F. Giustino, J. Phys.: Condens. Matter **24**, 202201 (2012).

[58]M. Landmann, E. Rauls, and W. G. Schmidt, J. Phys.: Condens. Matter **24**, 195503 (2012).

[59]L. Thulin and J. Guerra, Phys. Rev. B **77**, 195112 (2008).

[60] M. Mikami, S. Nakamura, O. Kitao, H. Arakawa, and X. Gonze, Jpn. J. Appl. Phys. Part 2 **39**, L847 (2000).

[61] M. Mattesini, J. S. de Almeida, L. Dubrovinsky, N. Dubrovinskaia, B. Johansson, and R. Ahuja, Phys. Rev. B **70**, 115101 (2004).

[62] X. G. Kong, Y. Yu, and T. Gao, Eur. Phys. J. B **76**, 365 (2010).

[63] M. Abbasnejad, E. Shojaee, M. R. Mohammadizadeh, M. Alaei, and R. Maezono, J. Appl. Phys. **100**, 261902 (2012)

[64] H. M. Lawler, J. J. Rehr, F. Vila, S. D. Dalosto, E. L. Shirley, and Z. H. Levine, Phys. Rev. B **78**, 205108 (2008).

[65] X. Gonze and et al., Z. Kristallogr. **220**, 558 (2005).

[66] X. Gonze and et al., Comput. Mater. Sci. **25**, 478 (2002).

[67] J. P. Perdew, K. Burke, and M. Ernzerhof, Phys. Rev. Lett. **77**, 3865 (1996).

[68] http://opium.sourceforge.net/

[69] N. Troullier and J. L. Martins, Phys. Rev. B **43**, 1993 (1991).

[70] H. B. Schlegel, J. Comp. Chem. **3**, 214 (1982).

[71] W. Hanke and L. J. Sham, Phys. Rev. B **21**, 4656 (1980).

[72] S. Albrecht, L. Reining, R. Del Sole, and G. Onida, Phys. Rev. Lett. **80**, 4510 (1998).

[73] M. Rohlfing and S. G. Louie, Phys. Rev. B **62**, 4927 (2000).

[74] P. Puschnig and C. Ambrosch-Draxl, Phys. Rev. B **66**, 165105 (2002).

[75] P. Lindan et al., Z. Kristallogr. **220**, 567 (2005).

[76] J. K. Burdett, T. Hughbanks, G. J. Miller, J. W. Richardson, and J. V. Smith, J. Am. Chem. Soc. **109**, 3639 (1987).

[77] R. Wyckoff, Crystal Structures, 2nd ed. (Interscience, New York, 1964).

[78] K. V. K. Rao, S. V. N. Naidu, and L. Iyengar, J. Am. Ceram. Soc. **53**, 124 (1970).

[79] V. W. H. Baur, Acta Cryst. 14, 214 (1961)

[80] T. A. Kandiel, L. Robben, A. Alkaim, and D. Bahnemann, Photochem. Photobiol. Sci. 12, 602 (2013).

[81] M. R. Ranade, A. Navrotsky, H. Z. Zhang, J. F. Banfield, S. H. Elder, A. Zaban, P. H. Borse, S. K. Kulkarni, G. S. Doran, and H. J. Whitfield, Proc. Natl. Acad. Sci. **99**, 6476 (2002).

[82] J. Zhu, Zach M. Curr. Opin. Colloid **14**, 260 (2009)

[83] T. E. Tiwald and M. Schubert, Proc. SPIE **4103**, 19 (2000).

[84] N. Hosaka, T. Sekiya, C. Satoko, and S. Kurita, J. Phys. Soc. Jpn. **66**, 877 (1997).

Table I. Theoretical electronic band gaps calculated by GW method reported in literatures. For rutile, "D" in the bracket means direct band gap energy and (I) in the bracket means indirect band gap energy.

|  | $E_{gap}^{GW}$ | | | | |
|---|---|---|---|---|---|
| Rutile | 3.34(I)/3.38(D)[a] | 3.59(D)[b] | 3.40(D)[c] | 3.46(D)[d] | |
| Anatase | 3.56[a] | 3.83[b] | 3.70[c] | 3.73[d] | 3.79[e] |
| Brookite | 4.45[d] | 3.68[f] | | | |
| Columbite | | Not given | | | |
| Baddeleyite | | Not given | | | |
| Cotunnite | | Not given | | | |
| Pyrite | | Not given | | | |
| Fluorite | 2.367(2.369)[g] | | | | |

[a] (LDA+GW) Ref. 55
[b] (PBE+GW) Ref. 56
[c] (DFT+U+GW) Ref. 57
[d] (PBE+GW) Ref. 58
[e] (PBE+GW) Ref. 59
[f] (HSE06+GW) Ref. 58
[g] (PBE+GW) Ref. 62

Table II. Space group, lattice constants (in Å) and angles (in degree), multiplicities and Wyckoff letters of nonequivalent atoms and the corresponding relative atomic coordinates for $TiO_2$ polymorphs. Available experimental data for lattice parameters are also listed for comparison.

| $TiO_2$ Structure | Space group | Structure parameters | Wyckoff positions |
|---|---|---|---|
| Rutile | $P4_2/mnm$ (136) | a=b=4.641, 4.587[a], 4.593[b], 4.594[c] <br> c=2.968, 2.954[a], 2.959[b], 2.958[c] | Ti 2a (0, 0, 0) <br> O 4f (0.305, 0.305, 0) |
| Anatase | $I4_1/amd$ (141) | a=b=3.797, 3.782[a], 3.785[b,d] <br> c=9.720, 9.502[a], 9.512[b], 9.514[d] | Ti 4a (0, 0, 0) <br> O 8e (0, 0, 0.206) |
| Brookite | Pbca (61) | a=9.263, 9.184[e] <br> b=5.510, 5.447[e] <br> c=5.167, 5.145[e] | Ti 8c (0.129, 0.089, 0.862) <br> O1 8c (0.010, 0.148, 0.182) <br> O2 8c (0.229, 0.108, 0.536) |
| Columbite | Pbcn (60) | a=4.581, 4.541[f], <br> b=5.578, 5.493[f] <br> c=4.921, 4.906[f] | Ti 4c (0, 0.178, 1/4) <br> O 8d (0.271, 0.380, 0.419) |
| Baddeleyite | $P2_1/c$ (14) | a=4.855, b=4.906, c=5.104 <br> β=100.24 | Ti 4e (0.275, 0.058, 0.217) <br> O1 4e (0.060, 0.318, 0.356) <br> O2 4e (0.450, 0.758, 0.455) |
| Cotunnite | Pnma (62) | a=5.219, b=3.354, c=6.871 | Ti 4c (0.265, 0.250, 0.080) <br> O1 4c (0.373, 0.250, 0.396) <br> O2 4c (-0.018, 0.750, 0.359) |
| Pyrite | Pa-3 (205) | a=b=c=4.894 | Ti 4a (0, 0, 0) <br> O 8c (0.340, 0.340, 0.340) |
| Fluorite | Fm-3m (225) | a=b=c=4.827 <br> (conventional cubic cell) | Ti 4a (0, 0, 0) <br> O 8c (1/4, 1/4, 1/4) |
| Tridymite | $P6_3/mmc$ (194) | a=b=5.938, c=9.683 | Ti 4f (1/3, 2/3, 0.437) <br> O1 2c (1/3, 2/3, 1/4) <br> O2 6g (1/2, 1/2, 0) |

[a] 15 K Ref. 76
[b] 295 K Ref. 76
[c] 298 K Ref. 77
[d] 301 K Ref. 78
[e] Ref. 79
[f] Ref. 12

Table III Band energies (in eV) at special *k*-points in the first Brillouin zones of TiO$_2$ polymorphs calculated by DFT-GGA and GW methods. The energies of valence band maximum and conduction band minimum are indicated by bold font. The valence band maximum from DFT-GGA calculation is set to 0 eV.

|   | DFT-GGA | | GW | |   | DFT-GGA | | GW | |   | DFT-GGA | | GW | |
|---|---|---|---|---|---|---|---|---|---|---|---|---|---|---|
|   | Ev | Ec | Ev | Ec |   | Ev | Ec | Ev | Ec |   | Ev | Ec | Ev | Ec |
|   | Rutile | | | |   | Anatase | | | |   | Brookite | | | |
| Z | -1.38 | 2.83 | -1.34 | 4.46 | Z | -0.22 | 2.17 | -0.23 | 3.75 | Γ | **0** | **2.38** | 0.35 | **4.21** |
| A | -0.66 | 2.49 | -0.52 | 4.08 | A | -0.25 | 3.39 | -0.27 | 5.15 | Z | -0.18 | 2.64 | 0.15 | 4.55 |
| M | -1.02 | 1.83 | -0.91 | 3.49 | M | -0.06 | 3.16 | -0.08 | 4.92 | T | -0.27 | 2.70 | 0.04 | 4.63 |
| Γ | **0** | **1.80** | 0.24 | 3.47 | Γ | -0.10 | **2.08** | -0.10 | **3.63** | Y | -0.16 | 2.46 | 0.20 | 4.33 |
| R | -0.97 | 1.84 | -0.86 | **3.42** | R | -0.64 | 3.15 | -0.73 | 4.81 | S | -0.47 | 2.81 | -0.17 | 4.79 |
| X | -0.70 | 2.45 | -0.56 | 4.07 | X | -0.83 | 3.29 | -0.98 | 5.00 | X | -0.56 | 2.70 | -0.21 | 4.67 |
|   |   |   |   |   | Δ | **0** | 3.04 | ***-0.02*** |   | U | -0.66 | 2.85 | -0.37 | 4.78 |
|   |   |   |   |   |   |   |   |   |   | R | -0.51 | 2.89 | -0.20 | 4.83 |
|   | Columbite | | | |   | Baddeleyite | | | |   | Cotunnite | | | |
| Γ | **0** | 2.63 | **0.31** | 4.46 | Z | -0.43 | 2.54 | -0.08 | 4.40 | Γ | -0.17 | **1.78** | 0.21 | **3.22** |
| Z | -0.28 | **2.59** | -0.02 | **4.40** | Γ | **0** | 2.34 | 0.39 | 4.18 | Z | -0.91 | 1.96 | -0.64 | 3.45 |
| T | -0.50 | 2.93 | -0.26 | 4.84 | Y | -0.33 | 2.75 | 0.03 | 4.59 | T | -0.35 | 2.10 | 0 | 3.74 |
| Y | -0.23 | 3.08 | 0.09 | 4.95 | A | -0.79 | 2.60 | -0.46 | 4.48 | Y | **0** | 1.93 | **0.33** | 3.44 |
| S | -0.33 | 2.96 | -0.05 | 4.89 | B | -0.15 | **2.33** | 0.21 | **4.08** | S | -0.70 | 2.37 | -0.40 | 3.95 |
| X | -0.11 | 2.72 | 0.19 | 4.60 | D | -0.68 | **2.33** | -0.36 | 4.13 | X | -0.20 | 2.14 | 0.17 | 3.72 |
| U | -0.40 | 2.84 | -0.14 | 4.70 | E | -0.92 | 2.52 | -0.64 | 4.29 | U | -0.33 | 2.11 | -0.02 | 3.63 |
| R | -0.78 | 2.99 | -0.54 | 4.85 | C | -0.90 | 2.63 | -0.55 | 4.49 | R | -0.84 | 2.37 | -0.54 | 3.74 |
|   | Pyrite | | | |   | Fluorite | | | |   | Tridymite | | | |
| X | -0.26 | 2.09 | 0.24 | 3.86 | W | -0.23 | 1.80 | 0.01 | 3.33 | Γ | **0** | 3.22 | **-0.02** | 5.65 |
| R | -1.29 | **1.39** | -0.96 | 3.06 | L | -2.47 | 1.42 | -2.44 | 2.93 | A | **0** | 3.48 | **-0.02** | 5.90 |
| M | -0.88 | 2.01 | -0.41 | 3.71 | Γ | -0.85 | **1.09** | -0.69 | **2.54** | H | -0.46 | 4.07 | -0.61 | 6.47 |
| Γ | **0** | 1.68 | 0.51 | 3.39 | X | **0** | 1.76 | **0.28** | 3.33 | K | -0.14 | 3.65 | -0.20 | 6.06 |
|   |   |   |   |   | K | -0.43 | 1.80 | -0.19 | 3.24 | M | **0** | 3.63 | **-0.04** | 6.04 |
|   |   |   |   |   |   |   |   |   |   | L | -0.25 | 3.76 | -0.34 | 6.15 |

Table IV. Electronic band gaps (in eV) for TiO$_2$ polymorphs with rutile, anatase, brookite, columbite, baddeleyite, pyrite, fluorite, cotunnite, and tridymite structures calculated by DFT-GGA and GW methods.

|  | $E_{gap}^{DFT}$ | $E_{gap}^{GW}$ |
|---|---|---|
| Rutile | 1.80 (D) | 3.18 (I) |
| Anatase | 2.08 (I) | 3.71 (I) |
| Brookite | 2.38 (D) | 3.86 (D) |
| Columbite | 2.59 (I) | 4.09 (I) |
| Baddeleyite | 2.33 (I) | 4.08 (I) |
| Cotunnite | 1.78 (I) | 2.89 (I) |
| Pyrite | 1.39 (I) | 2.55 (I) |
| Fluorite | 1.09 (I) | 2.26 (I) |
| Tridymite | 3.22 (D) | 5.67 (D) |

Table V. Calculated macroscopic static dielectric constants of TiO$_2$ polymorphs.

| Rutile | Anatase | Brookite | Columbite | Baddeleyite | Cotunnite | Pyrite | Fluorite | Tridymite |
|---|---|---|---|---|---|---|---|---|
| 5.71 (E⊥c) | 5.08 (E⊥c) | 5.31 (E//a) | 6.59 (E//a) | 6.13 (E//a) | 8.50 (E//a) | 8.08 | 9.56 | 2.39 (E⊥c) |
|  |  | 4.28 E//b) | 4.79 (E//b) | 6.31 (E//b) | 8.14 (E//b) |  |  |  |
| 7.33 (E//c) | 4.83 (E//c) | 4.40 (E//c) | 6.03 (E//c) | 5.57 (E//c) | 6.54 (E//c) |  |  | 2.40 (E//c) |

Figure captions

FIG. 1. Polyhedra structures for the TiO$_2$ polymorphs: (a) rutile, (b) anatase, (c) brookite, (d) columbite, (e) baddeleyite, (f) cotunnite, (g) pyrite, (h) fluorite, and (i) tridymite. Ti and O atoms are represented by big blue and small red spheres respectively.

FIG.2 Phonon density of states for tridymite-structured TiO$_2$. The inset shows the enlarged figure from -3.0 THz to 3.0 THz.

FIG. 3. Enthalpies (in eV, for four TiO$_2$ formula units) of TiO$_2$ polymorphs under different hydrostatic pressure calculated by DFT-GGA. The inset shows the enthalpy between 0 GPa and 11 GPa.

FIG. 4. The band structure and the corresponding DOS of (a) rutile, (b) anatase, (c) brookite, and (d) columbite calculated by DFT-GGA. Yellow points indicate the values obtained with the GW method. The valence band maximum from the DFT-GGA calculation is set to 0 eV.

FIG. 5. The band structure and the corresponding DOS calculated by DFT-GGA for four TiO$_2$ polymorphs that has been proposed in high-pressure studies: (a) baddeleyite, (b) cotunnite, (c) pyrite, and (d) fluorite. Yellow points indicate the values obtained with the GW method. The valence band maximum is set to 0 eV.

FIG. 6. Electronic states and optical absorption property of a newly proposed tridymite-structured TiO$_2$: (a) the band structure and the corresponding DOS calculated by DFT-GGA and the values obtained with the GW method are indicated by the yellow points, (b) imaginary part of the complex dielectric function calculated by solving the Bethe-Salpeter equation.

FIG. 7. Polarization-dependent real and imaginary parts of the complex dielectric function calculated by solving the Bethe-Salpeter equation for mineral TiO$_2$ polymorphs: (a) rutile, (b) anatase, (c) brookite, and (d) columbite. The experimental dielectric function (ref. 83 for rutile and ref. 84 for anatase) are shown for comparison.

FIG. 8. Absorption coefficients for (a) rutile, (b) anatase, and (c) brookite. The experimental dielectric function (ref. 83 for rutile and ref. 84 for anatase), the experimental absorption coefficients (ref. 38 for rutile and ref. 37 for anatase, and brookite) are shown for comparison.

FIG. 9. Imaginary parts of the complex dielectric function calculated by solving the Bethe-Salpeter equation for TiO$_2$ polymorphs with baddeleyite, cotunnite, pyrite, and fluorite structures. A tick at 3.1 eV is specially added to discern the optical transitions in the visible light range.

FIG. 1

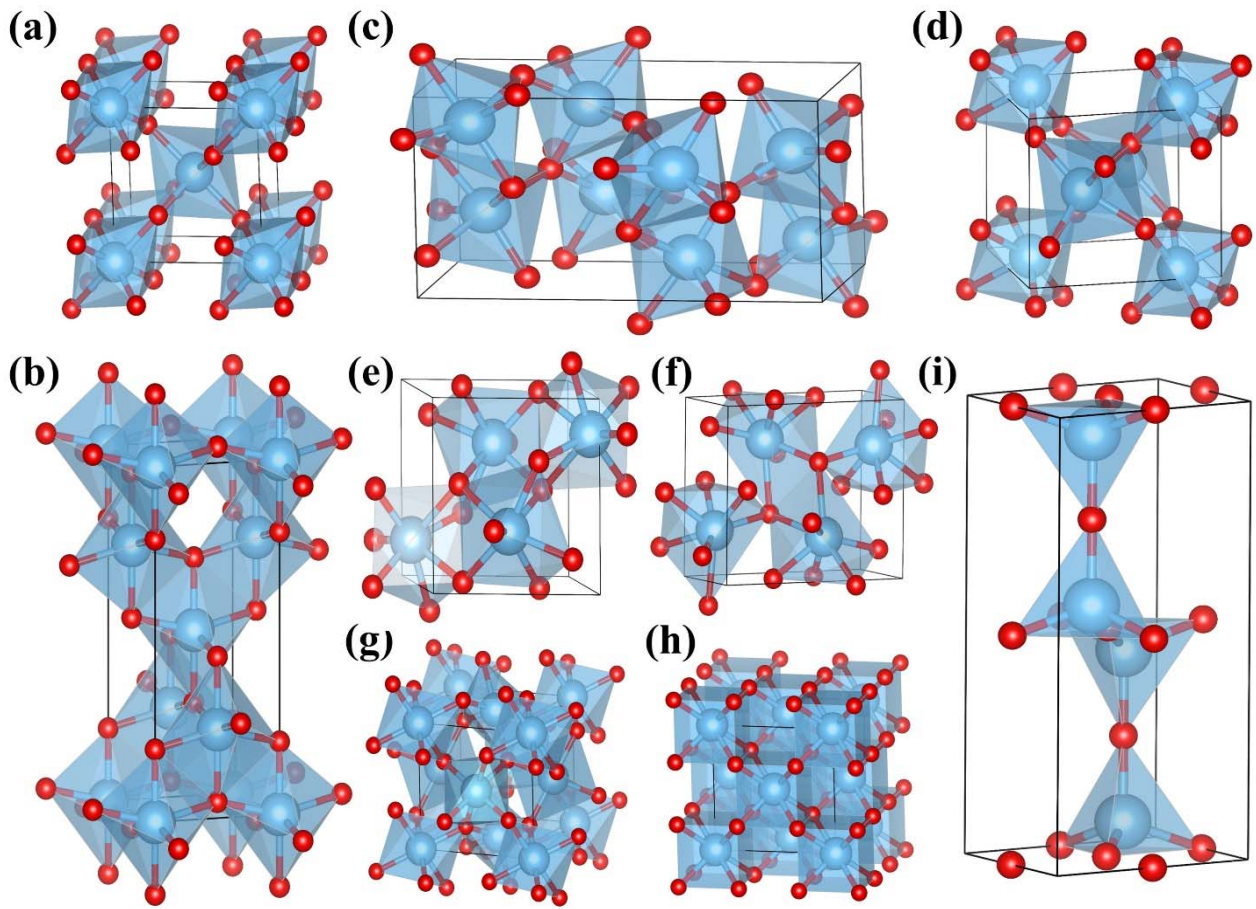

FIG.2

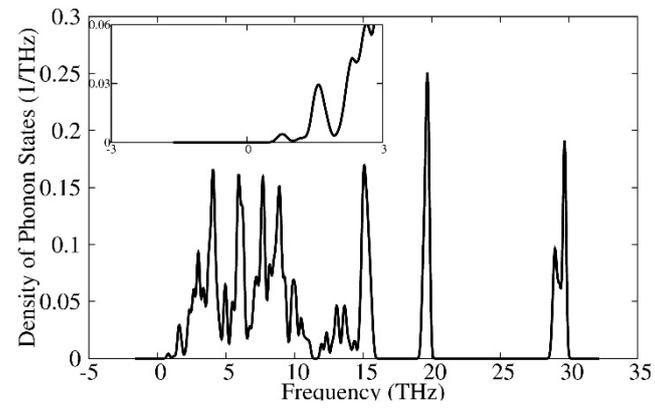

FIG.3

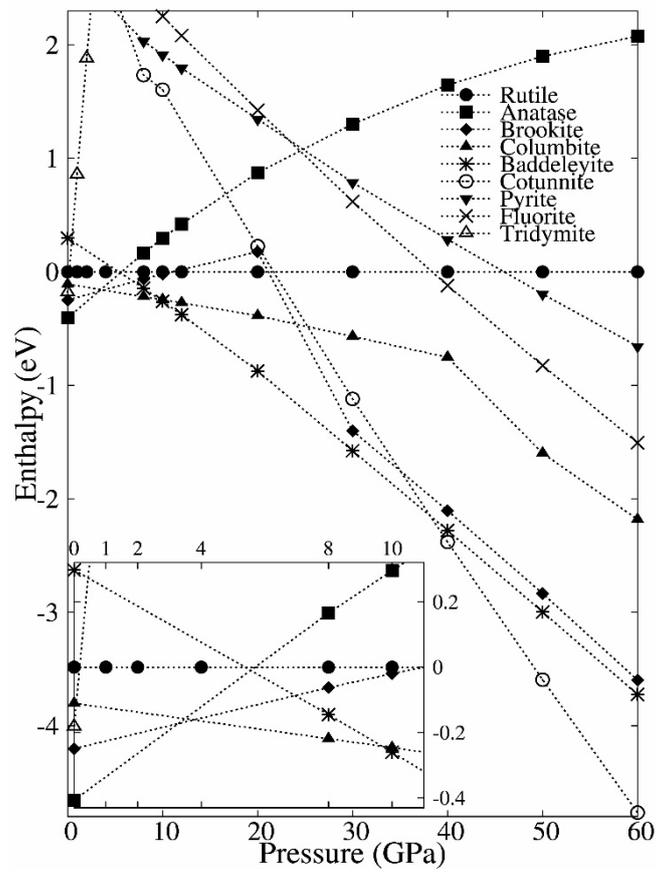

FIG. 4

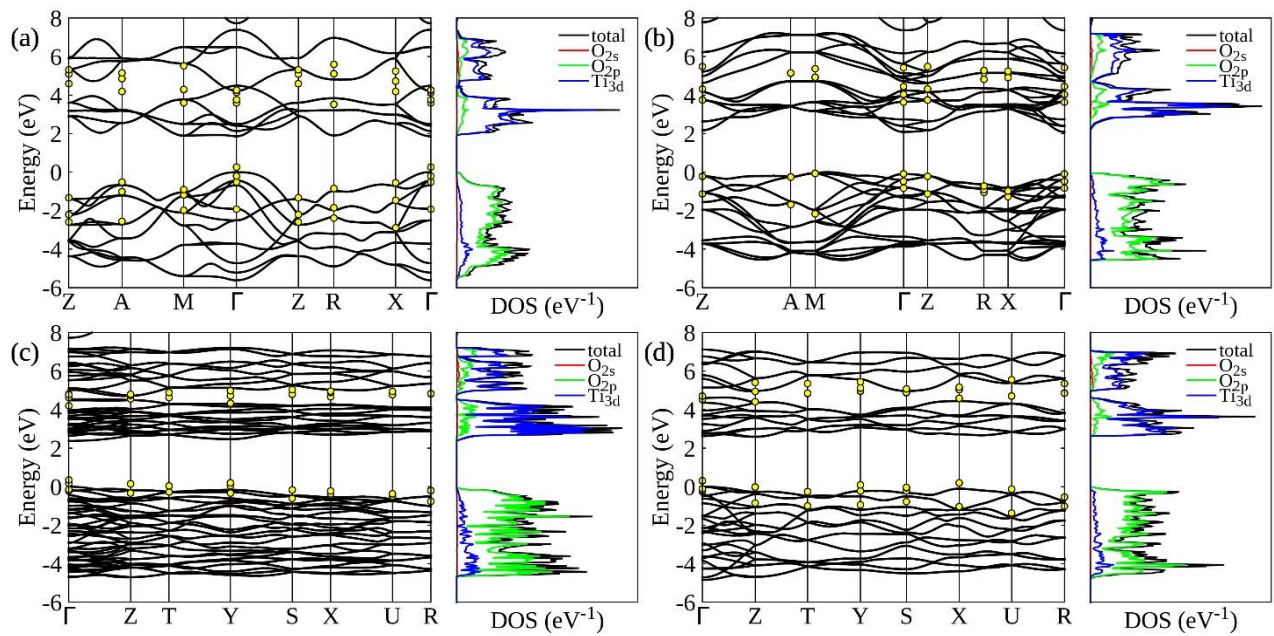

FIG. 5

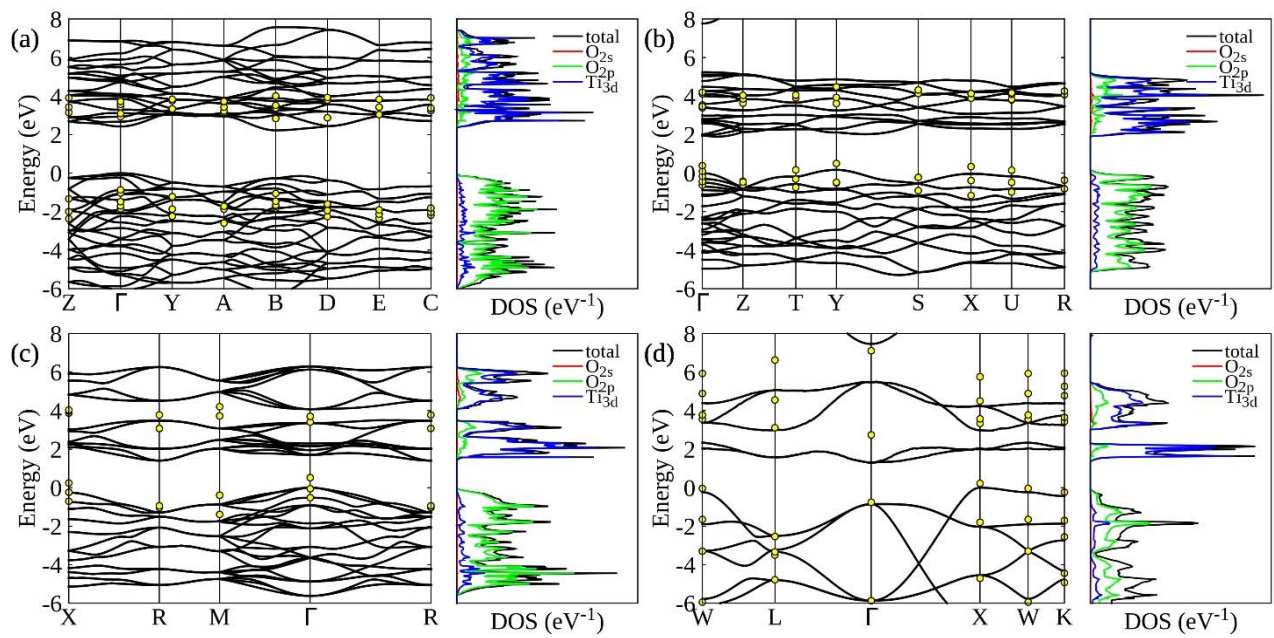

FIG. 6

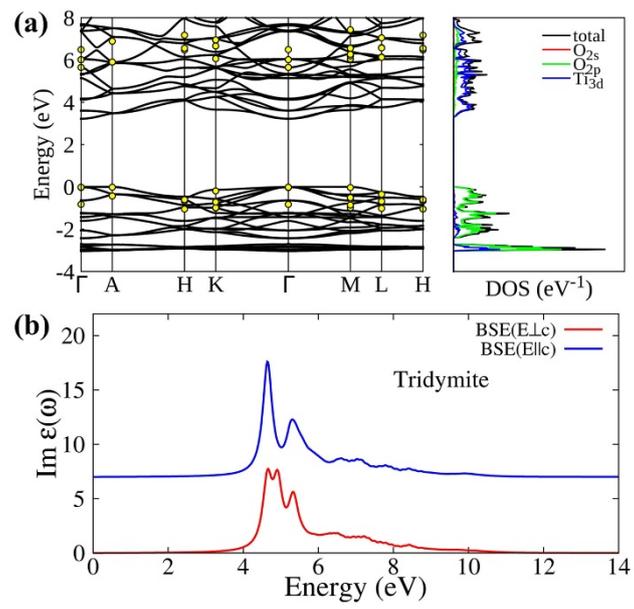

FIG. 7

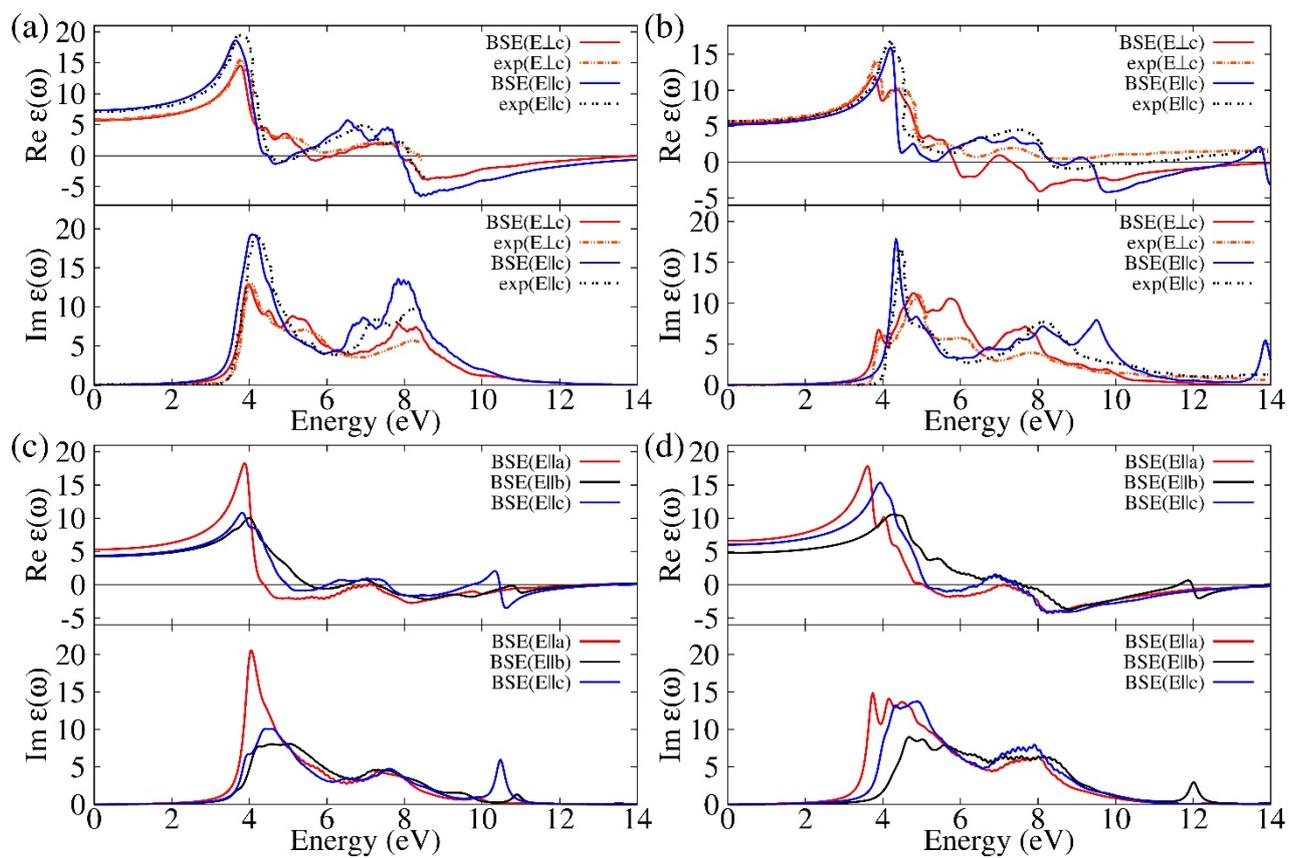

FIG. 8

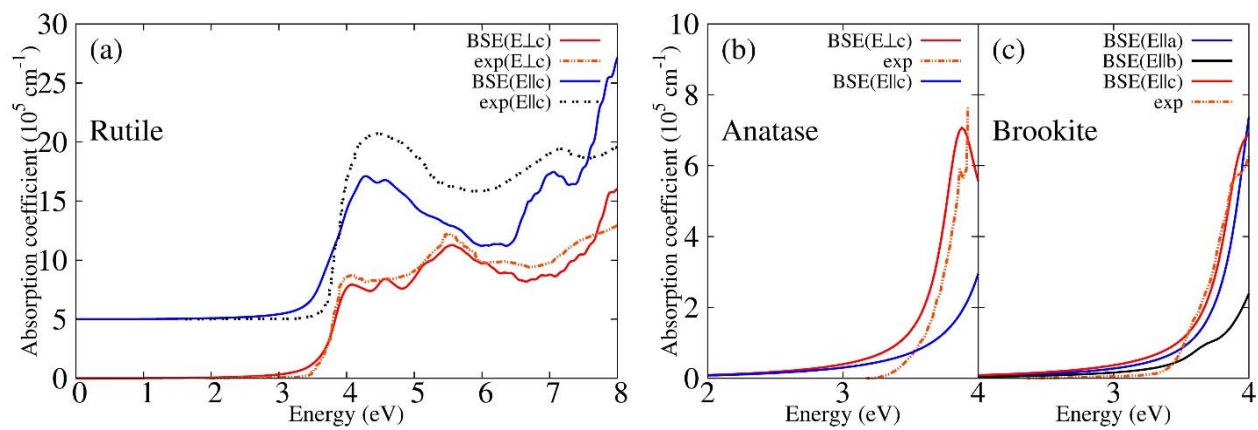

FIG. 9

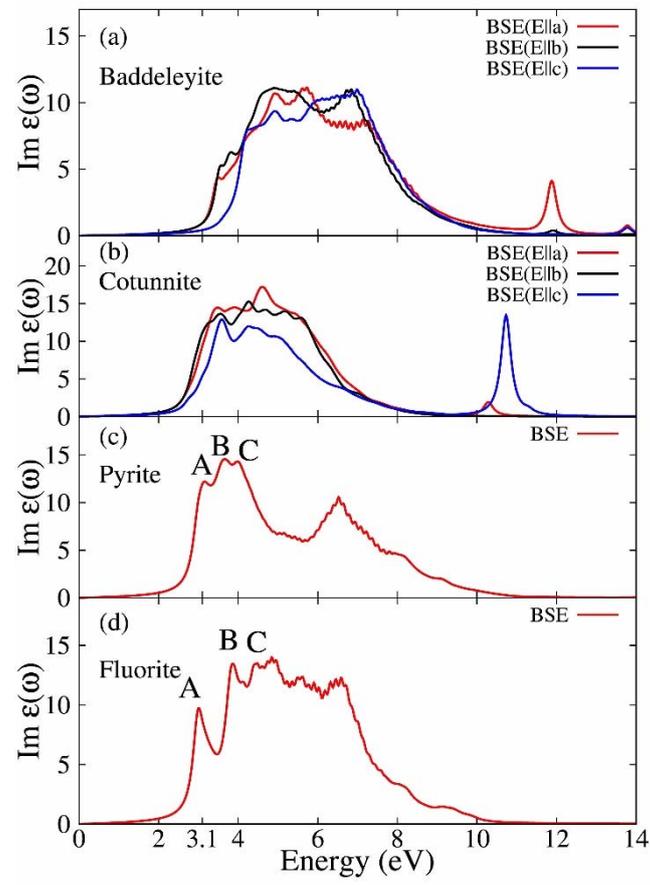